# Global Prompt Proton Sensor Network: Monitoring Solar Energetic Protons based on GPS Satellite Constellation

Yue Chen[1], Steven K. Morley[1], Matthew R. Carver[1]

[1]Space Science and Applications Group, Los Alamos National Laboratory, Los Alamos, New Mexico, USA

**Key Points:**

- Energetic particle instruments carried by GPS satellites form a powerful global sensor network to monitor solar energetic proton events
- GPS observations characterize solar energetic protons by features including origins, distributions and dynamic cutoff L-shell values
- Long-term GPS data can greatly advance understanding and forecasting solar proton events together with other satellite missions

**Abstract.** Energetic particle instruments on board GPS satellites form a powerful global prompt proton sensor network (GPPSn) that provides an unprecedented opportunity to monitor and characterize solar energetic protons targeting the Earth. The medium-Earth-orbits of the GPS constellation have the unique advantage of allowing solar energetic protons to be simultaneously measured from multiple points in both open- and closed-field line regions. Examining two example intervals of solar proton events, we showcase in this study how GPS proton data are prepared, calibrated and utilized to reveal important features of solar protons, including their source, acceleration/scattering by interplanetary shocks, the relative position of Earth when impinged by these shocks, the shape of solar particle fronts, the access of solar protons inside the dynamic geomagnetic field, as well temporally-varying proton distributions in both energy and space. By comparing to Van Allen Probes data, GPS proton observations are further demonstrated not only to be useful for qualitatively monitoring the dynamics of solar protons, but also for quantitative scientific research including determining cutoff L-shells. Our results establish that this GPPSn can join forces with other existing solar proton monitors and contribute to observing, warning, understanding and ultimately forecasting the incoming solar energetic proton events.

# 1. Introduction

Solar explosive events, such as flares and coronal mass ejections (CMEs), frequently emit particles accelerated to extremely high speeds. These so-called solar energetic particles (SEPs), including electrons, protons, alpha particles, and heavier ions (also called HZE ions) with energies ranging from ~10s keV up to GeV, zip across the heliosphere with some of them heading towards the Earth. When ground- and/or space-based instruments register significantly higher solar particle fluxes compared to the background level, we say an SEP event is observed arriving at the Earth.

SEP events with high intensities and long durations pose a major space radiation hazard associated with both dose (ionizing and non-ionizing) and single event effects. For example, protons with energy > 50 MeV can easily penetrate though nominal spacecraft shielding (Jiggens et al., 2018), and thus endanger astronauts (e.g., Sanzari et al., 2014) and/or electronic systems on board (e.g., Tylka et al., 1996), particularly for those in high-altitude or interplanetary orbits with weak or no protection from the geomagnetic field. Also, particles bombarding the upper atmosphere in major SEP events generate air showers of induced secondary particles, e.g., neutrons, which can also be dangerous for passengers on board airplanes over the Polar Regions (Wilson et al., 2003). During extreme and rare solar cosmic ray events that contain solar protons with energies exceeding GeV, induced energetic neutrons (> 1 – 10s MeV) can even reach the ground level at mid-latitude (Shea and Smart, 2000) and may cause problems in both avionics and sea-level microelectronic systems through single event upset or latchup (e.g., Dyer et al., 2006 and Cellere et al., 2008), depending on the neutron fluence and the sensitivity of exposed systems. Therefore, monitoring, warning, simulating, understanding, and forecasting SEP events have practical significance for communities including space and aerospace.

Naturally, a prerequisite for reliably predicting SEP events is to specify the morphology and underlying physics of those solar particles, which include their origin, acceleration and propagation processes inside the heliosphere, as well as their energy spectrum, duration, and access to low altitudes (i.e., penetration depths) inside the dynamic geomagnetic field upon their arrival at the Earth.  SEP events can be roughly separated into two categories based on observations: impulsive events that are electron-rich with high $^3$He/$^4$He ratios and often accompanied by type III radio bursts, and gradual events that are proton-rich and often associated with type II radio emissions (and also type III sometimes; see, e.g., Desai and Giacalone, 2016 and references therein). Type III bursts in impulsive events are believed to generate from outward steaming electrons (Bastion et al., 1998) and particles in those events often have sources with limited temporal (~ hours) and spatial distributions inside the heliosphere. In comparison, solar energetic protons accelerated by CME-related shocks dominate gradual events, of which the major ones have longer durations (~ days) and wider spatial range (> tens of degrees in heliospheric longitude) (Cane et al., 1988). More comprehensive discussions on SEPs can be found in the reviews by Reames (2013a,b) as well as Desai and Giacalone (2016), and this study will focus on those solar proton events (SPEs).

Currently, SPEs are routinely monitored by operational satellite missions from NASA and National Oceanic and Atmospheric Administration (NOAA), including the solar wind monitors at the upstream Lagrangian 1 point of Sun-Earth system as well as Geostationary Operational Environmental Satellites (GOES) in the geosynchronous (GEO) orbit. NOAA's Polar Operational Environmental Satellites (POES) constellation in low-Earth-orbits (LEO) also contribute to SPE observations when they travel across the high-latitude region. In addition, science exploration missions such as NASA's Van Allen Probes mission, previously known as RBSP (Mauk et al., 2013), have also made observations of energetic protons over the last several years. Predictions of SPEs have recently achieved progress (e.g., Winter and Ledbetter, 2015), but are still beyond our reach in the foreseeable future; therefore, observational studies including more available high-quality data of SPEs should be increasingly useful.

In this work, we showcase that the long-term, continuous proton observations from the Global Positioning System (GPS) satellite constellation can be utilized for monitoring and understanding SPEs. Indeed, the GPS constellation has a long history of carrying energetic particle instruments developed by Los Alamos National Laboratory (LANL) since 1980s (Cayton et al., 1992). Originally geared toward nuclear-test ban treaty verification, these instruments have previously demonstrated their uses in scientific research on the natural space radiation environment, including both trapped electrons (e.g., Reeves et al., 2003, Chen et al. 2007, Morley et al., 2010, and Morley et al., 2016) and protons (Chen et al., 2016) inside the radiation belts. Recently, LANL has made decades of GPS particle data available to the public (Morley et al., 2017). These data are indexed at data.gov, and can be retrieved directly from https://www.ngdc.noaa.gov/stp/space-weather/satellite-data/satellite-systems/gps/. This data release was part of the coordinated efforts of enhancing national space-weather preparedness following the instruction of the National Space Weather Strategy (National Science and Technology Council, 2015a) and National Space Weather Action Plan (National Science and Technology Council, 2015b). While the initial public release included some proton measurements (see Morley et al., 2017), the science quality of these data was further improved and demonstrated by Carver et al. (2018) which cross-calibrated GPS proton measurements with GOES data. The latest public release of GPS energetic particle data has been updated to include cross-calibrated proton measurements.

Availability and readiness of GPS proton data is the essential first step, while appropriately using those data for physics revealing studies is as critical, if not more. This work aims to illustrate how GPS proton data can be used to advance our understanding of the SPE morphology by carefully examining two SPE intervals in March 2012 and September 2017. Indeed, as a constellation in medium-Earth-orbits (MEOs), multiple GPS satellites provide simultaneous observations covering both the open- and closed-field line regions, which makes GPS data a unique resource for SPE observations, complementing observations from other orbits. Section 2 briefly summarizes the GPS proton data used for this study, and detailed analyses of those data across a 15-day interval in March 2012 are presented in Section 3. In Section 4, GPS and RBSP proton fluxes are compared, as well as the identified cutoff L-shell values in the September 2017 SPE interval. This work is concluded in Section 5 with a summary of our findings and possible future directions.

**2. Overview of GPS Proton Data**

Starting from 2000, among the > 24 GPS satellites, there are usually more than 10 of them carrying charged particle instruments developed by LANL. The latest instrument is the Combined X-ray Dosimeter (CXD), replacing the previous version of Burst Detector Dosimeter for Block II-R (BDD-IIR, Cayton et al., 1998). A detailed description of the CXD instrument is given by Tuszewski et al. (2004), and we briefly recap here. Besides measuring electrons (see Morley et al., 2016), the CXD instrument also measures protons with energies from ~6 MeV up to greater than 75 MeV within five channels using three sensors: the low-energy particle sensor containing two proton channels 6–10 MeV (P1) and 10–50 MeV (P2), the first high-energy X-ray and particle detector with one proton channel 16–128 MeV (P3), and the second high-energy X-ray and particle detector with two proton channels 57–75 MeV (P4) and >75 MeV (P5). These above energies are just nominal definitions, while the latest response functions for those proton channels have been provided by Carver et al. (2018), which also shows differential and integral proton fluxes can be derived by using a forward model and fitting to the observed counts of all five proton channels simultaneously. That work also derived and applied calibration factors (< ~2) to improve the consistency with the proton fluxes measured by the GOES Energetic Particle Sensor instrument (Hanser, 2011). Readers are referred to Carver et al. (2018) for further details of the generation of CXD proton fluxes. Preliminary work (e.g., Cayton et al., 2007 and Morley et al., 2017) has shown the promise of GPS data for measuring solar energetic protons and understanding their penetration in the magnetosphere.

GPS data used in this study are differential proton fluxes with a time resolution of 4 min. These fluxes cover an energy range of 10 - $10^4$ MeV in the original data files, while in this study proton fluxes at > 200 MeV are not used, and we discuss further on this in Section 3.2. An overview plot of GPS proton flux distributions is provided in Figure 1 for the first selected 15-day interval in March 2012. In the beginning ~6 days, GPS data are dominated by background signals, while orbital variations are clearly visible after the arrival of the first major SPE at ~148 hr. Basically, low fluxes (the green strips) during the SPE indicate the region at low L-shells that is well protected by the geomagnetic field, and the high flux region (the yellow and red areas) at large L-shells that solar protons can easily access. In the first SPE, two interplanetary (IP) shocks are observed to arrive at Earth on 07$^{th}$ March 2012 0347 UT (S1) and 08$^{th}$ 1053 UT (S2), according to the IP shock list on

http://umtof.umd.edu/pm/figs.html, identified from both the SOHO proton monitor (Ipavich et al., 1998) as well as the upstream solar wind speeds in Figure 1N. Hereinafter we call the event from ~148 hr to 288 hr as SPE1. The second proton event starting from ~ 307 hr is called SPE2, which is associated with a flare occurring on 13$^{th}$ March 1712 UT at (N19, W66) on the Sun's surface, as observed in GOES X-ray data (see the NASA GOES major SPE list at https://cdaw.gsfc.nasa.gov/CME_list/sepe/).

As discussed, satellites of the GPS constellation orbit in circular MEOs in six different planes with altitudes of ~ 20,200 km, having four operating satellites in each orbit plane (sometimes an extra satellite for backup). These satellites distribute almost evenly around the Earth, and thus provide simultaneous coverage on multiple L-shells (Panel K), magnetic local times (MLT in Panel L), and latitudes/hemispheres (Panel M). Their nominal 55º inclination and 20,200 km altitude allow the GPS satellites to spend a significant portion of time in both closed- and open-field line regions, as in Panel K. Note these satellites cannot access L-shells below ~ 4, which determines the minimum cutoff L-shell observable to GPS. Here we loosely define the region with $L > 10$ (<10) as open- (closed-) field line region. GPS data with their high temporal and spatial resolutions, allow us to observe SPEs in unprecedented details, as will be presented in the next Section.

It is interesting to observe how the arrivals of SPEs and IP shocks relate to geomagnetic storms. For instance, in Panel N, SPE1 as well as the following shock S1 arrives during a prestorm phase, while the arrival of S2 occurs during the late recovery phase and precedes a major geomagnetic storm. Indeed, the storm sudden commencement on 08$^{th}$ 1103 UT caused by S2 associated with an Interplanetary CME (ICME) is identified in the Richard-Cane ICME list at http://www.srl.caltech.edu/ACE/ASC/DATA/level3/icmetable2.htm. The arrival of SPE2 happens in the late recovery phase of a moderate storm. Also, the arrival of S1 and S2 can be identified from solar wind speeds—a gradual increase of speeds for S1 and a sudden increment for S2, while the arrival of SPE2 shows no such shock signature, implying possibly different acceleration/transport processes experienced by solar protons of the two events. Figure 1 demonstrates that GPS proton data can at least be used for monitoring and warning SPEs qualitatively.

**3. Analyzing GPS Observations of two SPEs in March 2012**

In this section, we present details of how to examine, calibrate and analyze GPS data for understanding solar protons quantitatively, using the 15-day period shown in Figure 1 as an example. First, GPS proton observations from the ten satellites are replotted as normalized flux distributions in Figure 2, aiming to highlight the relative increases of protons across different energies. Proton fluxes at each energy from all satellites during the first day (1$^{st}$ March) of the interval are averaged and used for the normalization. As in most panels from A to J, it is seen that for SPE1 before ~192 hr, protons with high energies at ~100 MeV have higher increment factors, while more significantly, high energy protons arrive—the appearance of green color—earlier than low energy protons in the term of flux ratios. Visually, during the decay of SPE1, ns60 data show that lower energy protons start to decay earlier than higher energy, which is very different from other satellites, and thus are deemed unreliable and excluded from the following analyses.

Panels K to M also present how proton flux ratios evolve with time at three different locations as well as their energy dependence. It can be seen that the flux ratios at L ~ 4.4 are generally lower than high L regions, particularly at decay times. Also, at times ~168 +/- 12 hr between the arrivals of two shocks, it is noticeable that flux ratios at L ~ 4.4 decrease significantly, especially for the lower energy of 16 MeV, while those at higher L-shells remain almost constant. Both observations reflect that the penetration of protons varies by particle energy as well as the recovery of geomagnetic field as seen from the Dst index in Panel N. The flux-ratio ranges at any given moment in Panels K to M are partially due to the variation range of L-shells (i.e., dL). Specifically in Panel M, another reason for the high variations of 16 MeV proton fluxes observed between ~216 – 276 hr is the presence of a transient residual proton belt, which will be further presented and discussed in Section 3.3. Next, we analyze SPE characteristics for different physics, starting with the use of single satellite observations with original high time resolution, to multi-satellite observations, and finally to binned observations with continuous spatial coverage.

## 3.1 Examining Velocity Dispersion from Single-satellite Observations with High Time Resolution

The "velocity dispersion" of solar protons—that particles arrive sequentially in inverse velocity order—has been widely reported by previous works and used to help constrain the acceleration process (e.g., Tylka et al., 2003, Reames, 2009, Kouloumvakos et al., 2015 and Ding el al., 2016). This feature is also observable in GPS data for both the gradual SPE1 and impulsive SPE2 as shown in Figure 2. To take a closer look, we selected a 24 hr period in the beginning of the each SPE and replotted proton distributions in Figure 3. Data used here have the original 4 min time resolution. In Panel B1 for SPE1, when ns61 travels inside the open-field line region at time ~149 - 151 hr, it is clear that the highest energy protons arrive first followed sequentially by lower energies. Likewise, in Panel B2 for SPE2, at ~306 - 307 hr with ns61 inside the open-field line region, similar velocity dispersion can be identified although this one has shorter time lags compared to SPE1.

These velocity dispersion can be linearly fitted to qualitatively determine the expected solar particle release times (e.g., Tylka et al., 2003 and Reames, 2009), which are plotted in Figure 4 for both SPE1 and SPE2. We start our discussion with SPE2. Flux ratio curves for different proton energies are plotted in Panel A2, in which a threshold ratio factor $f_c = 4$ is selected to identify the "arrival" times of protons. The value of 4 is chosen because the exact onset times of flux increases are not always clearly observed for all energies; therefore, we look for a clearly-identifiable increase above the background level. Panel B2 plots the identified arrival time of each energy as a function of the inverse velocity normalized to the light speed, and the data points can be well fitted by a straight line. Slope of the fitted line represents the magnetic path length traveled along by solar protons from the source, which has a value of 0.99 AU for SPE2, and the vertical intercept gives the time $T_0$ of 305.93 hr. Considering that the arrival times of $f_c$ lag by ~0.3 hr relative to the real onset times as in Panel A2, the real solar particle release time for SPE2 should be ~305.6 hr, which is ~17.6 UT on 13$^{th}$ March. This time falls well within the flare time range starting from 17.2 UT and flux peak arriving at 17.7 UT as observed by GOES X-ray measurements from the NASA major SPE list. Since the CXD instruments do not resolve proton pitch angles mainly due to the absence of in-situ magnetic field measurements, here for SPE2 we assumed a ~0º pitch angle for solar protons relative to the interplanetary magnetic field, an assumption often used (e.g., Reames 2009) for impulsive events when particles are accelerated close to the Sun. Thus the source location for the impulsive SPE2 can be clearly identified assuming no significant CME-related proton scattering (accelerating) involved. This is consistent with no observation of shock and related CME during the event (see, e.g., the lists mentioned in Section 2 as well as the SOHO CME list at https://cdaw.gsfc.nasa.gov/CME_list/). Note on this day of each year the distance between the Sun and Earth is slightly below the average of 1 AU, thus the calculated path length suggest that protons in SPE2 may travel along a Parker spiral with a very small curvature.

Similarly, the path length and solar particle release time for SPE1 are calculated as shown in Panel B1. Here we assumed a large pitch angle of ~80º relative to the interplanetary magnetic field to have a path length of 1.6 AU. Note particles' pitch-angle cosine is proportional to the path length but has no effect in determining $T_0$. Considering the arrival times of $f_c$ lag by ~1.1 hr to the real onset times as in Panel A1, the real solar particle release time on the Sun for SPE1 should be ~144.0 hr, which is ~0 UT on 07$^{th}$ March. This time is ahead of the solar flare starting at 0.2 UT on the day as observed by GOES X-ray measurements. Therefore, different from SPE2, it is likely that solar protons in SPE1 have experienced significant CME scattering en-route from Sun to the Earth, which may explain why here the calculated release time differs from the observed flare time. Ding et al. (2016) have studied the same SPE using STEREO observations located at very different heliospheric longitudes, and they have reached a similar conclusion on the involvement of a CME. Since time scales for the dispersed arrival of solar protons are less than ~ 1 hr, as shown in Figure 4, this velocity dispersion phenomenon can be studied by using GPS data with the original high time resolution of 4 min.

**3.2 Large-scale Pictures of SPEs Derived from Multi-satellite Observations with High Time Resolution**

The difference between arrival times of the IP shock and energetic protons also reveals the relative position of the Earth meeting the shock front. According to Cane et al. (1988), given the strongest acceleration (i.e., the main source of solar energetic protons) occurring near the "nose" of a shock moving radially outward from the Sun, the arrivals of shocks compared to the time profiles of protons in gradual SPEs tell us where the Earth locates inside the shock front in term of solar longitude: when Earth is at solar longitudes to the east (west) of the source, proton intensities peak before (after) the shock arrival considering how the magnetic field lines connecting to the source sweep across the Earth. As in the left panels of Figure 5, the shock S1 arrives well before the peak of proton intensities and thus the Earth was first hit by the very west "flank" of the shock front in SPE1. The same is also be true for the second shock S2 since it arrives more than ~10 hr prior the peak of proton intensities at ~182 hr as in Figure 2K-2M. However, in the impulsive SPE2, there is no IP shock but the flare F3 related to the onset of proton intensities as in the right panels of Figure 5. The ~1 hr delay of proton onsets after F3 is consistent with observations in Figure 4.

Data in Figure 5 include observations from multiple GPS satellites to improve the continuity of temporal coverage. In Panels A1, B1, and C1 for SPE1, there is no significant time difference between flux onset times at different L-shells, which can be explained by that the spatial scale of the particle front is much larger than the size of GPS constellation (~ 8.5 Earth radii across). This is consistent with the large spatial scales usually observed for gradual SPEs. By comparison, for SPE2 in the right panels, the flux onset times in the closed-field line region (Panels B2 and C2) seem to be ~0.5 hr earlier than that in open-field line region (Panel A2). This may indicate the shape of the particle front has a small scale in the heliospheric z-direction, for instance, a tongue or more-or-less pointed front shape. We will return to this later.

Spreads of the normalized fluxes in Figure 5 are noticeable, which reminds us of the importance of proper intra-calibration between different GPS instruments. One such study was carried out and the results are presented in Figure 6. First, we did a crude calibration which compares the averaged proton flux spectrum observed from each single GPS satellite within the selected L-shell and time ranges. Panel A1 first compares flux spectra measured during the three quiet days at L-shells between 4 and 5. Of more interest is to compare proton flux spectra measured over 20 hr during SPE1 as in Panel B1, which shows more consistent measurements than quiet times except for the curve from ns63 (in red) being an outlier. Therefore observations from satellite ns63 are excluded from the following studies. Also, since proton fluxes from different satellites spread out at energies > 200 MeV, particularly those for ns62 (in yellow), we confine our analysis to proton energies below 125 MeV in this study. This energy value is consistent with the energy ranges of GPS integral fluxes that were used quantitatively in the initial cross-calibration with GOES by Carver et al. (2018). More careful calibrations were also done by comparing fluxes measured within the same 3-hr-long time, 1-hr-long MLT, and 0.2-wide L-shell bins. Calibration results at L-shells within 4 and 5 are shown in Panels A2, including data points from quiet, SPE1 and SPE2 periods. Also results inside the open-field line region are shown in Panel B2. Since almost all data points in the two panels are confined within the two dashes lines, we conclude that, for proton flux distributions sourced from multiple GPS satellites, only differences with flux ratios larger than 3 are significant enough to be counted as real.

Figure 7 exhibits several snapshots of global proton distributions observed by GPS satellites. For SPE1 as shown in Figure 7a, 100 MeV protons have first increments observed at almost all L-shells at 147.792 hr as in panel A1, then quickly fill the belt region as in B1 and C1; while 16 MeV proton increments start ~15 min later as in Panel B2 than 100 MeV protons, which is consistent with the late arrival as discussed in Section 3.1. As in Figure 7b for SPE2, Panel A1 clearly shows 100 MeV proton increments start only at L-shells within 8, e.g., comparing the low flux of ns55 (ns57) to the high flux of ns59 (ns56) in Northern (Southern) hemisphere, and then the high fluxes expand to higher L-shells as seen in Panels B1 and C1. Similarly, 16 MeV proton increments start latter as in Panel B2 only at L-shells within ~5-8, and then expand to higher L shells at the next moment (Panel C2). The L-dependent arrival of solar protons may be due to their spatial distribution in the heliospheric z-direction as above discussions on Figure 5, while the lack of significant presence of 16 MeV protons at L < ~5 is expected by their limited penetration inside the geomagnetic field and consistent with previous studies (e.g., Qin et al., 2019 and references therein).

**3.3 Solar Proton Distributions and Cutoff L-shells from Binned Observations**

Another conventional analysis method is to average data from multiple GPS satellites in bins of L-shell and time, so as to have more continuous temporal and spatial coverage, while the trade-off is that small-scale features are averaged out (e.g., Hudson et al. 1998 and Morley et al., 2017). Two such examples are shown in the top two panels in Figure 8 for the time-varying flux distributions of 16 and 100 MeV protons, respectively, as a function of L-shell. Here the arrival of protons can be clearly seen, while the time lags of low energy protons are still recognizable for SPE1 but not identifiable at all for SPE2 due to the averaging over 3-hr bins. Also, during the late phase of SPE1 after ~ 216 hr, penetration depths of protons gradually withdraw due to the recovery of geomagnetic field (Leske et al., 2001), with residual 16 MeV protons at L-shells < 4.5 being visible in Panel A1. These residual protons, having a temporally moving distribution gap below the penetrating solar protons, explain the widely spreading flux distributions of 16 MeV protons at L ~ 4.4 +/- 0.4 between ~216 – 276 hr as plotted in Figure 2M. High flux levels of low-energy protons at the largest L = 10 in Panel A1 are seen to sustain longer than high-energy protons in Panel B1 during both SPE1 and SPE2.

Snapshots of flux distributions in L, MLT and energy space are also presented for three selected moments in Figure 8. Panels C1, D1 and E1 are for 16 MeV protons with low, medium and high flux levels during the SPEs respectively, and C2 – E2 are for 100 MeV protons. Proton distributions as functions of energy and L-shell in the Northern hemisphere are also plotted in Panels C3 – E3 for the same three moments. Note that the apparent L-MLT features in both C1 and C2 at 148.5 hr are mostly due to the 3-hr window width, which includes GPS observations both before and after the arrival of solar protons (e.g., see Panels A2 and B2 in Figure 7a), and thus are artifacts of this analysis method. That fewer solar protons can access lower L-shells can be seen in both Panels E1 and E2, and how the penetration depths depend on proton energies is also depicted in Panels C3 - E3. Even with similar flux levels for ~10s MeV protons in open-field line region, their penetration depths are very different comparing Panels D3 and E3.

In this study, we determined solar protons' penetration depth using the ratios between fluxes in the closed- and open-field line regions, analogous to previous work done by others (e.g., Leske et al., 2001). Examples at three selected moments are shown in Figure 9, with one row for each induvial time point. Panels in the left column present proton fluxes as a function of energy inside the closed-field line region (also called belt region), and the panels in the central column are proton flux distributions inside the open-field line region. Fluxes in the open-field line region are averaged for each individual energy, and are used to normalize fluxes inside the belt region. On the right, flux ratio distributions are shown for the three times and cutoff L-shells are identified when the flux ratio is below 0.2, which is the threshold ratio value used in this work. In Panel A3, solar protons are able to penetrate inside the minimum L-shell (~4) of the GPS orbits, while in Panel B3, cutoff L-shells can be identified from observations, while the presence of residual low-energy protons trapped at L < ~4.5 smear the picture to some degree (in other words, identifying cutoff L-shells is sensitive to the selected threshold ratio value for low-energy protons at this moment). The black curve plots proton's cutoff energies at different L-shells, calculated from Störmer's equation (assuming a static dipole magnetic field; Störmer, 1955), and the white curve plots the cutoff L-shells calculated by the Dst-dependent empirical relation developed by Ogliore et al. (2001) and Leske et al. (2001). Cutoff L-shells are clearly determined at the time of 307.5 hr in Panel C3—not sensitive to the selected threshold ratio—at slightly higher L than the ones from the Ogliore and Leske formula but with a similar shape. Cutoff L-shells determined by Störmer's equation systematically have substantially larger values. Employing the method described above, we are able to continuously determine cutoff L-shells for any given proton energy based upon binned GPS observations. Figure 10 shows the results for two energies derived for the 15-day interval in 2012. Basically, due to the minimum L ~ 4 for the GPS orbits, cutoff L-shells are mostly determined during the recovery phase of geomagnetic storms but unlikely in the main phases when solar protons often penetrate below the minimum L-shell of GPS, particularly for high energy protons as in Panel F. Cutoff L-shells based on GPS data are close to those determined using the Ogliore and Leske formula when Dst values are not much below zero, but not so between ~ 200 – 240 hr when Dst values are more negative, particularly for 100 MeV protons. Obviously, GPS observations can be used for monitoring solar protons penetrating into the magnetosphere, and cutoff L-shells derived from the GPS data set can be invaluable for exploring the statistical dependence(s) of protons' cutoff on energy and geomagnetic activities.

**4. Discussions**

One main question regarding the GPS proton data is how representative are those fluxes observed off-equatorially, considering the high-inclined orbits of this constellation. Based on the previous inter-calibration with GOES data at GEO (Carver et al., 2018), here we further compare GPS observations to those from the Relativistic Proton Spectrometer (RPS, Mazur et al., 2012) instrument carried by the Van Allen Probes mission that covers L-shells between ~1.2 – 6. RPS measures angular flux distributions of protons ranging from ~60 – 2000 MeV from the near equatorial plane within GEO. Here we chose to compare observations over two days—10$^{th}$ - 11$^{th}$ September 2017—as shown in Figure 11.

As an overview, Panels A and B plot proton fluxes observed by GPS ns72 and RBSP-b satellites, respectively, as a function of energy between 50-200 MeV over the two days. A major SPE occurs at ~16.5 UT on the first day 10$^{th}$ Sep. L-shell and MLT traces of both satellites are shown in Panels C and D, and two 5-hr periods were selected for comparison: 07-12 hr for quiet time and 31-36 hr for SPE time. Radial distributions of 100 MeV protons show large differences between ns72 and RBSP-b data during the quiet time (Panel E), indicating the different instrument backgrounds, while distributions at the SPE time agree with each other quite well in Panel F. It can be seen that the ratio factors between fluxes at L > ~ 5 are smaller than 2, slowly growing to ~5 as the L-shell decreases to < 4.5. Panel G shows the comparison as a function of energy, and the flux ratios between RBSP-b and ns72 have an average of 1.43 during the SPE time with a minimum of 0.39 at 200 MeV and a maximum of 2.16 at 100 MeV. A more complete picture is shown in Panel H for the SPE time, which shows GPS proton fluxes agree with RBSP-b data very well at L-shells outside of 5 (except for > ~180 MeV). For smaller L-shells, below 125 MeV within the energy range of our interest, RBSP-b fluxes are higher than ns72 values for ~100 MeV by a factor < 4, while RBSP-b fluxes are smaller for 10s MeV protons with a factor of ~ 0.25. A variety of reasons can contribute to these non-unity ratios, such as the different look directions of instruments, accuracy of detectors' response functions, fitting algorithms for differential fluxes, as well as the geomagnetic shielding being more significant at smaller L-shells. For the purposes of this work, these comparison factors for CXD fluxes are deemed generally acceptable for quantitative analysis of SPEs, particularly at L-shells > ~5, but the differences should be considered in future studies using observations from both missions.

Based on this favorable comparison, we use GPS proton data to determine solar protons' cutoff L-shells and compare them to those identified from observations by other missions. For this we chose another 14-day period in September 2017 as shown in Figure 12, in which intra-calibrated observations from 7 GPS satellites are binned the same way as in Figure 10. Two major SPEs are observed during this interval: the first event has an initial flux maximum for 100 (16) MeV protons appearing on the midday of 6$^{th}$ Sep. (beginning of 7$^{th}$ Sep.) followed by another maximum on early 8$^{th}$, and the second major SPE starting on 10$^{th}$ as in Panel E (A). The two flux maximums of the fist SPE may be caused by the forward IP shocks associated with two consecutive ICMEs (see Richard-Cane ICME list and Kasper/Stevens shock list at https://www.cfa.harvard.edu/shocks/wi_data/index.html), but detailed study on this relationship is beyond the scope of this work and thus is left to the future. Cutoff L-shells are identified for both 16 MeV (Panel C) and 100 MeV (Panel F) protons, particularly for the second major SPE during the late recovery phase of a geomagnetic storm (Panel G). Indeed, the second major SPE has been examined by O'Brien et al. (2018) which has cutoff L-shells determined from both RPS and NOAA Polar-orbiting Operational Environmental Satellite (POES) data. When using GPS proton fluxes for 63 MeV (one of the energies provided in the GPS data files) and a threshold flux ratio of 0.5 (same as the one in O'Brien et al. (2018)), our dynamically-derived cutoff L-shells track those derived using RPS 58 MeV protons very closely during this major SPE from 10$^{th}$ to 14$^{th}$ (not shown here). The GPS cutoff L-shells reside slightly inside of those from RPS with an averaged difference of ~ 0.3, and they almost overlap with those derived from POES 50 MeV protons. These agreements further demonstrate the scientific quality of the GPS proton data. The same SPE has also been examined by Luhmann et al (2018) and Qin et al. (2019) using multiple satellite data sets compared to simulations.

## 5. Summary and Conclusions

LANL energetic particle instruments on the GPS constellation have measured the natural space environment continuously over several decades. Due to their original purpose for national securities and other reasons, these data set has been under-used for scientific research. In 2017, LANL released GPS energetic particle data from 2000 through 2016 to public (Morley et al., 2017), and these data have been recently updated to include cross-calibrated energetic proton fluxes (Carver et al., 2018). There is a growing community awareness of the value of GPS particle measurements for space weather studies (e.g., Knipp and Giles, 2016, Olifer et al., 2018, and Wang et al., 2018), and we hope through this work that the role of GPS proton data can be better recognized by the space weather research community.

In summary, we have demonstrated that the LANL GPS energetic particle instruments form a powerful global prompt proton sensor network (GPPSn) that provides an unprecedented opportunity to monitor and study the characteristics of solar energetic protons arriving at Earth. The MEO orbits of this constellation allow solar energetic protons to be measured in both open- and closed-field line regions. Using two solar energetic proton event intervals as examples, we have examined the origin and possible effects of CME on the solar protons, as well as their spatial extends and energy distributions when they access the Earth's magnetosphere, by properly using GPS proton data in different ways. By comparing to Van Allen Probes data, we have also shown that GPS proton data can not only qualitatively monitor the dynamics of solar protons, but also be quantitatively valid for scientific researches including the analysis of geomagnetic cutoffs. Our results strongly suggest that this GPPSn can join force with other solar proton monitors and theoretical models, and contribute to the monitoring, warning, understanding and eventually prediction of incoming solar energetic proton events. Thus, this new copious data resource from the GPPSn has the great potential of enhancing our preparedness for severe space weather events in the future by enabling a broad range of new solar proton research.

**Acknowledgements** This work was performed under the auspices of the US Department of Energy and supported by the Laboratory Directed Research and Development (LDRD) program, award 20190262ER. We gratefully acknowledge the CXD instrument team at Los Alamos National Laboratory, as well as the RBSP RPS team for instrument development, providing measurements and allowing us to use their data. Thanks to CDAWeb for providing OMNI data. LANL GPS data used here are available from https://www.ngdc.noaa.gov/stp/space-weather/satellite-data/satellite-systems/gps/, and RBSP RPS data can be downloaded from http://mag.gmu.edu/ftp/RBSP/RPS/.

**Figures**

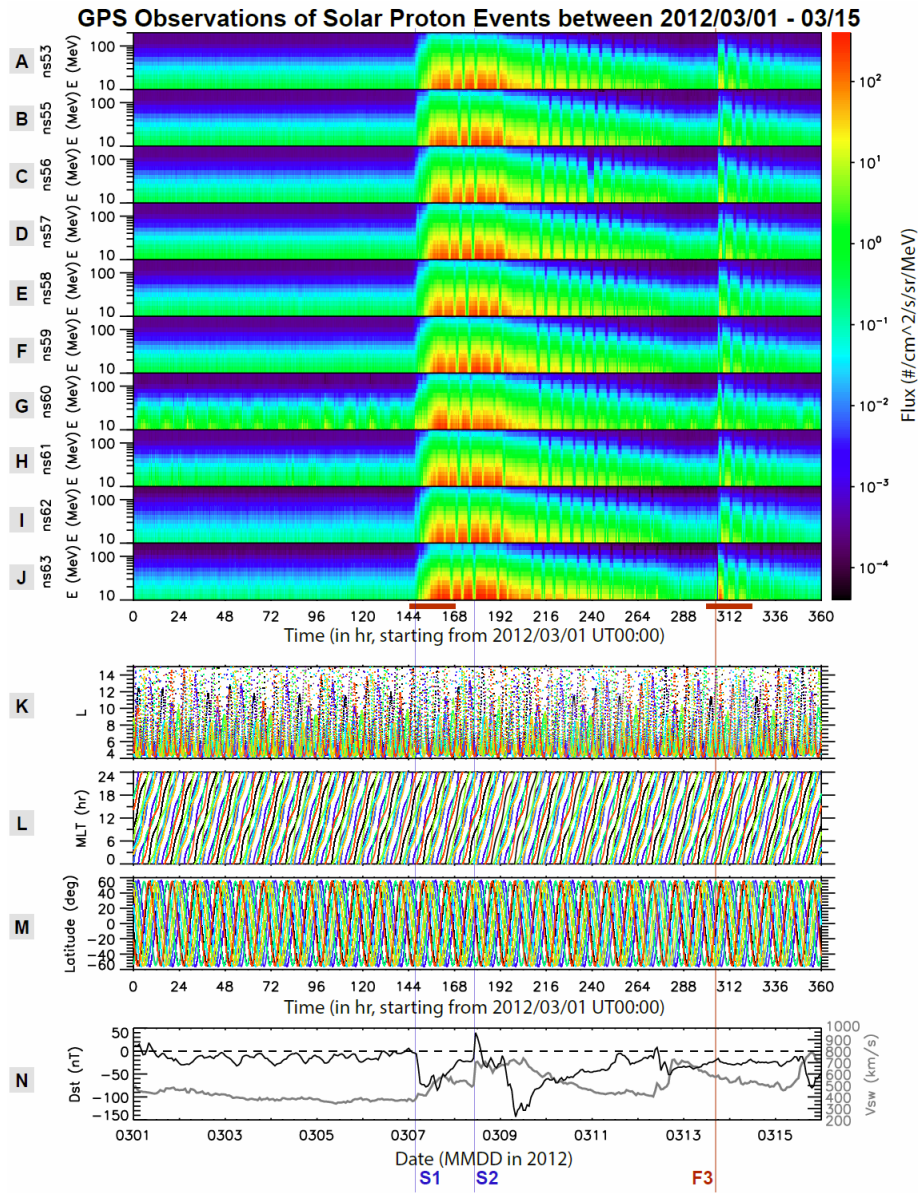

**Figure 1 Overview of original proton measurements from 10 GPS satellites during a 15-day interval in March 2012, including two solar proton events (SPEs). A to J)** Proton differential fluxes are presented as a function of energy and time for the 10 GPS satellites from ns53 to ns63, respectively. Here fluxes have a 4-min time resolution. A major gradual SPE is observed starting from the beginning of 07$^{th}$ March and the 2$^{nd}$ weaker SPE starting from the 13$^{th}$. Two red bars under Panel J are the 24 hr periods selected for further examining velocity dispersion as in Figure 3. Spatial coverages of related satellite orbits are displayed in **K)** for L-shells, **L)** magnetic local times (MLTs), and **M)** geographic latitudes in different color for each individual GPS satellite. **N)** Dst index (in black) and upstream solar wind speeds (gray). Arrivals of two interplanetary shocks S1 and S2 are marked out by blue vertical lines, as well as the associated solar flare (F3) in red.

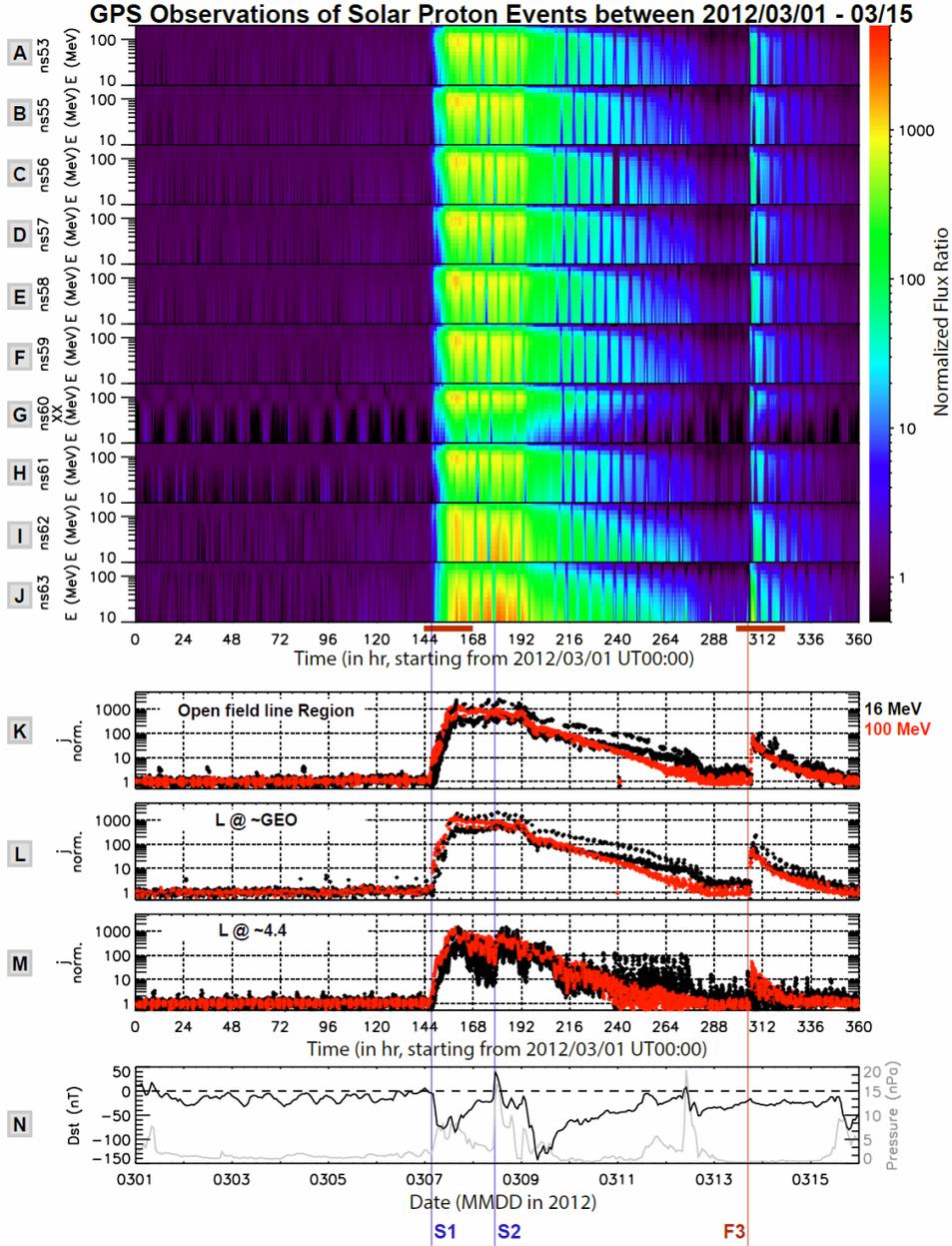

**Figure 2 Overview of normalized proton measurements from 10 GPS satellites during the same 15-day interval as in Figure 1. A to J)** Normalized proton differential fluxes are presented as a function of energy and time for the 10 GPS satellites. Proton fluxes at each energy point are normalized to the mean flux observed by all satellites during the first 24 hr of the interval. **K)** Temporal changes of normalized fluxes for 16 MeV (black) and 100 MeV protons in the open-field line region (L > 10). **L)** Normalized proton fluxes at ~GEO (L ~ 6.6, with variations within dL=+/-0.3). **M)** Normalized proton fluxes at L ~ 4.4 (+/- 0.4). **N)** Dst index (in black) and upstream solar wind pressure (gray). The same shocks S1 and S2 as well as the solar flare F3 are marked out by vertical lines. Two red bars under Panel J are the 24 hr periods selected for examining velocity dispersion as in Figure 3.

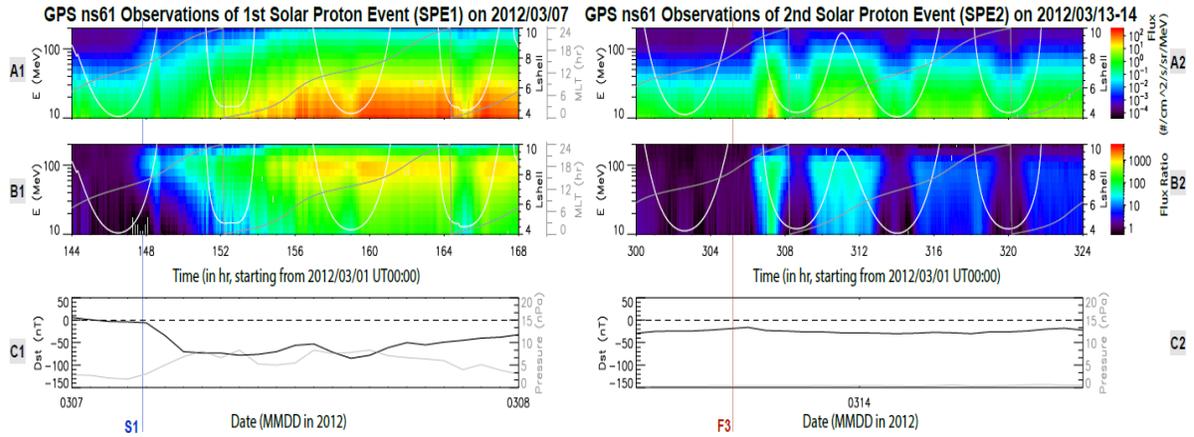

**Figure 3 Arrival of solar protons observed by a single GPS satellite (ns61) during the beginning of two SPEs.** Panels in left are for SPE1 and right for SPE2. **A1 & A2)** Proton differential fluxes from ns61 are presented as a function of energy over a 24 hr period for both SPE1 and SPE2. L-shells (in white) and MLT (gray) of the satellite are overplotted, respectively. **B1 & B2)** Normalized proton fluxes are presented in the same format. **C1 & C2)** Dst index (in black) and upstream solar wind pressure (gray). The arrival of S1 is marked out in left, as well as the solar flare F3 in right.

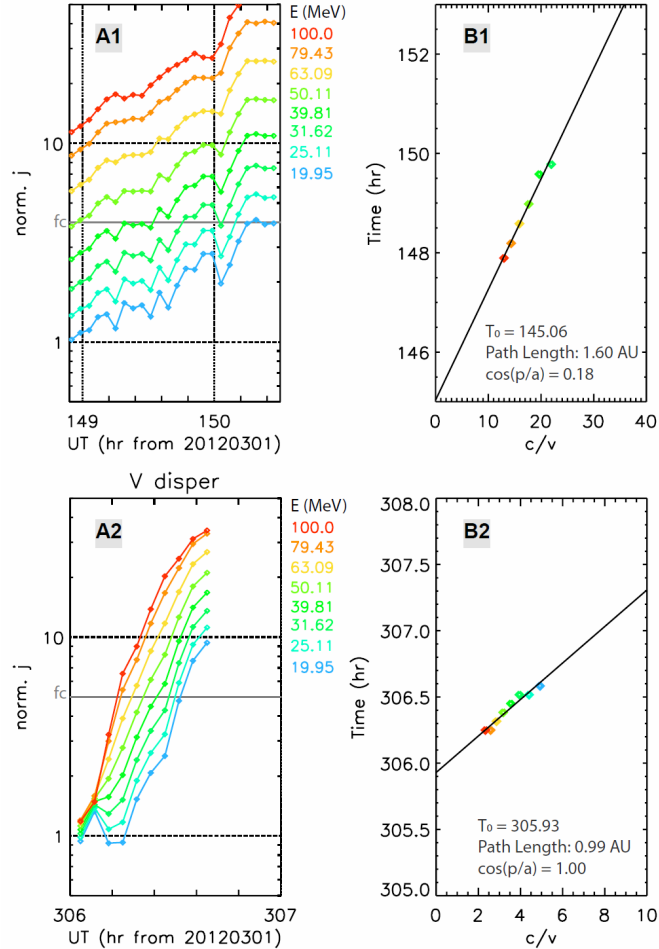

**Figure 4 Velocity dispersion of solar protons observed by ns61 in open-field line region during the two SPEs.** Panels in top are for SPE1 and bottom for SPE2. **A1 & A2**) Temporal changes of proton fluxes for different energies. The $f_c$ (= 4) is the flux threshold level selected to determine the arrival times of protons at different energies. **B1 & B2**) Arrival times of protons plotted as a function of the reciprocal of normalized speeds. The black straight line fits data points linearly. $T_0$ is the time derived for the related solar surface event, if any. Path length traveled by protons is given in the unit of AU. p/a is the assumed proton pitch angle.

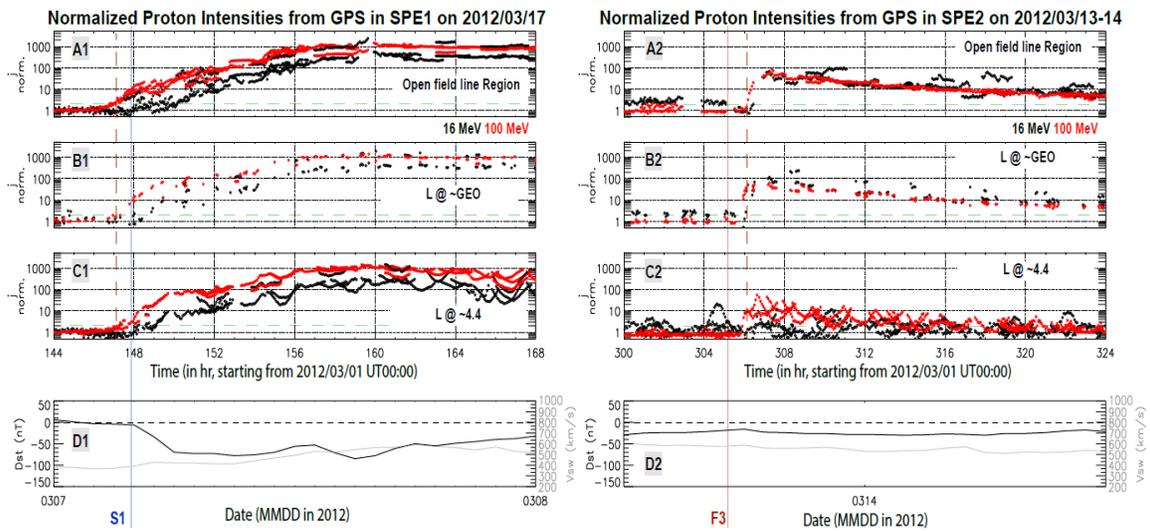

**Figure 5 Proton flux increments in-situ measured by GPS constellation at different L-shells during the beginning of two SPEs.** Panels in left (right) are for SPE1 (SPE2). **A1 & A2)** Normalized proton fluxes in the open-field line region during the same 24 hr as in Figure 3 for two energies: 16 (in black) and 100 (red) MeV. The dashed horizontal green lines indicate a ratio of 2, and the dashed vertical red lines mark the time for 100 MeV protons to go beyond that flux ratio observed in the open-field line region. **B1 & B2)** Normalized proton fluxes at ~ GEO. **C1 & C2)** Normalized proton fluxes at L ~ 4.4. **D1 & D2)** Dst index (in black) and upstream solar wind speeds (gray) over the 24-hr period. The arrival of S1 is observed during the period in left, as well as the solar flare F3 in right.

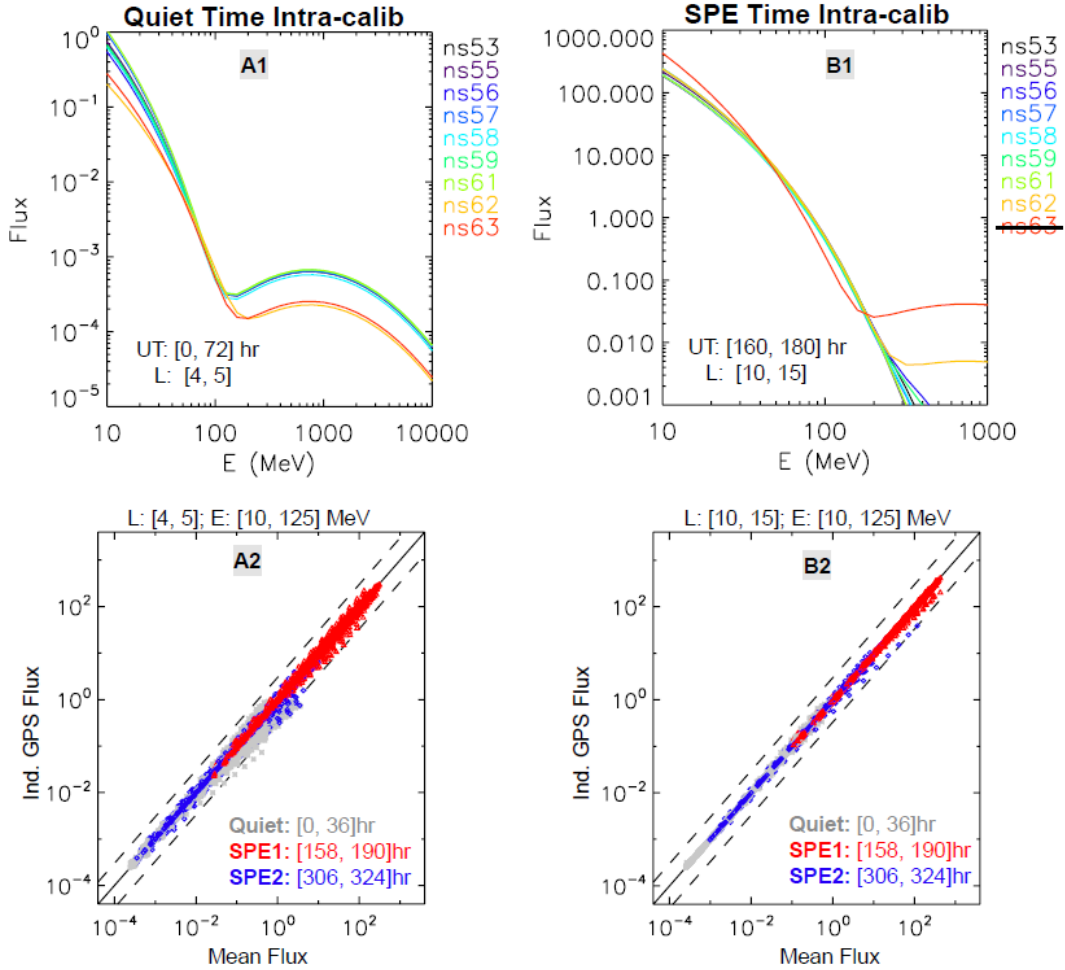

**Figure 6 Intra-calibrations of CXD proton measurements from different GPS satellites during the 15-day interval in March 2012. A1**) Averaged proton energy spectra at L-shells ranging between 4 and 5 during three quiet days. **B1**) Averaged proton energy spectra within the open-field line region during 20 peak hours of SPE1. **A2**) Proton fluxes from individual satellites compared to the mean fluxes from all GPS satellites within the same L, MLT and time bins. Data points are for L-shells between 4 and 5. Gray points for measurements during the quiet time, red for SPE1, and blue for SPE2. The solid diagonal line indicates perfect match, and the two dashed lines indicate ratio factors of 3. **B2**) Same as A2 but for data points within the open-field line region.

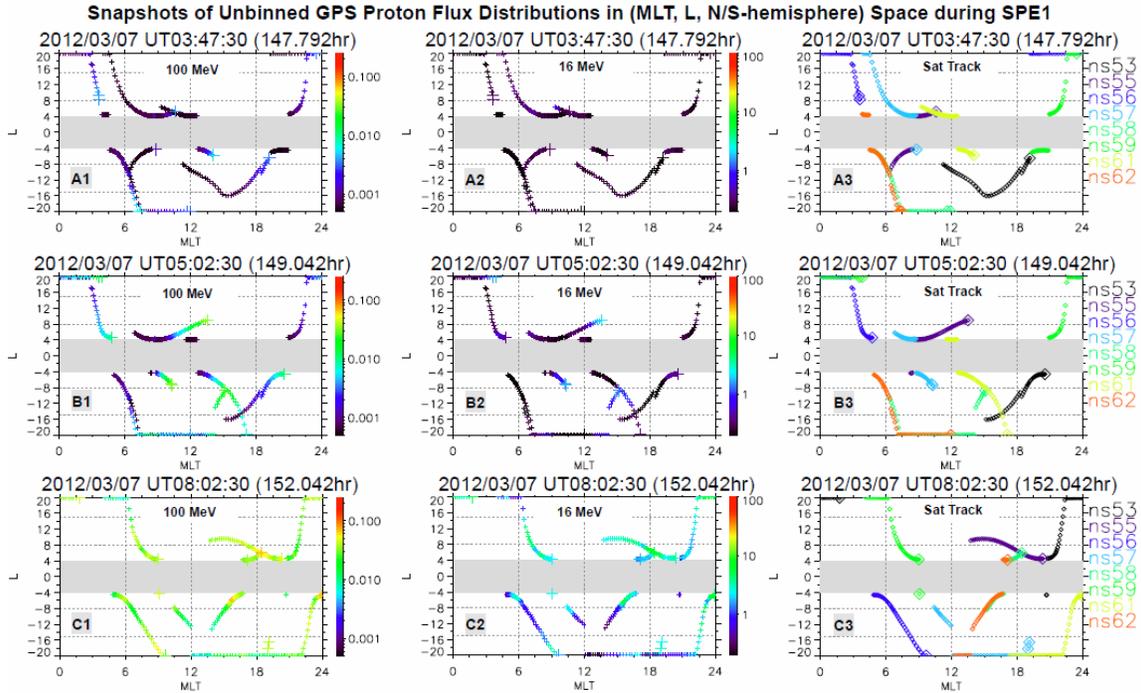

**Figure 7a Five-minute snapshots of proton spatial distributions in SPE1 as observed by GPS constellation.** Panels in three rows are for three moments at 147.792, 149.042 hr and 152.042 hr, respectively. Panels in left column are for flux distributions of 100 MeV protons inside the L-MLT space, and central for 16 MeV protons. Panels in right column show the 3-hr orbital tracks color-coded for individual satellites. Each data point (flux or orbital position) has a 4 min time resolution, with the large symbol(s) representing the data point(s) within the current 5 min window and other small one for those in the previous 3 hr. In each panel, positive (negative) L-shells are for orbits in the Northern (Southern) hemisphere, all data points at L > 20 are mapped down to L=20, and |L| < ~4.2 are shaded for the zone not accessible by GPS satellites.

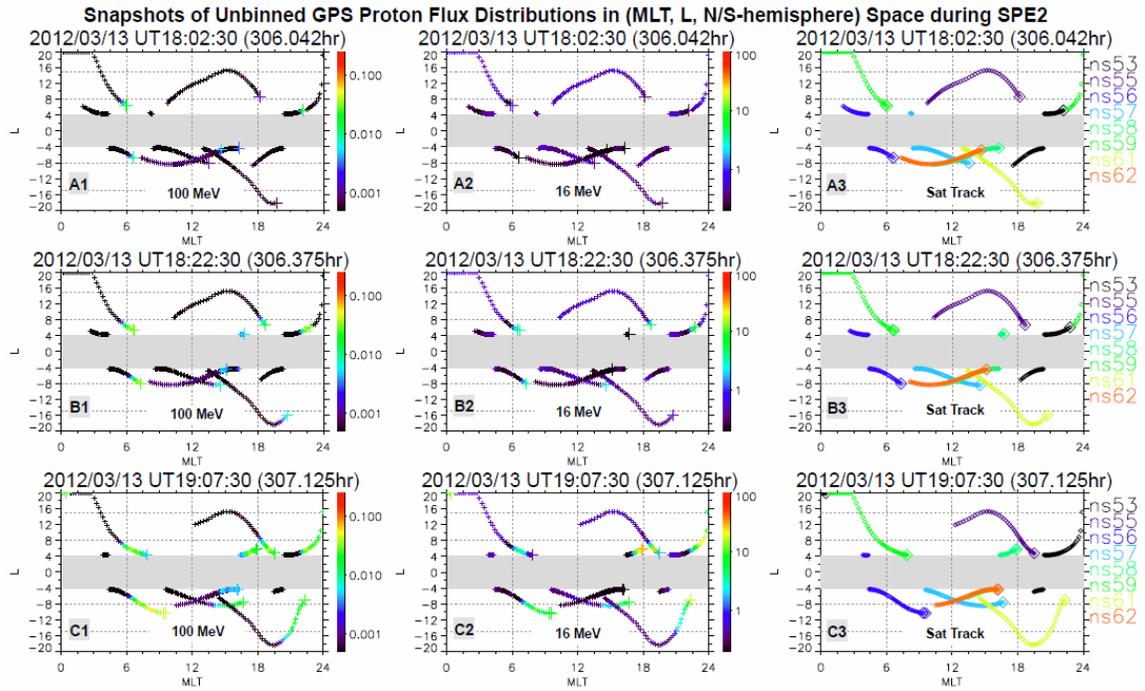

**Figure 7b Five-minute snapshots of proton spatial distributions in SPE2 as observed by GPS constellation.** In same format as Figure 7a.

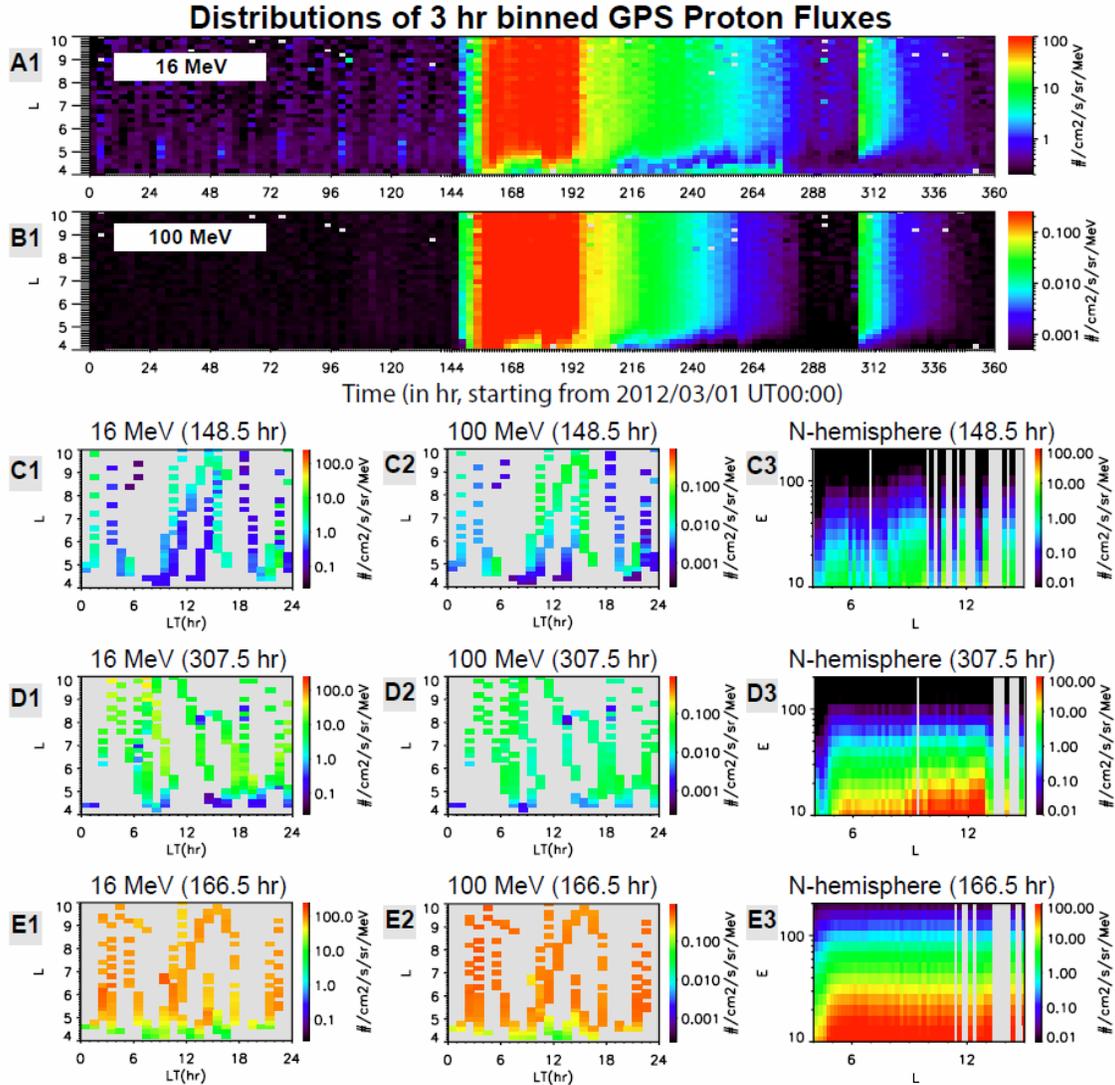

**Figure 8 Distributions of 3 hourly binned GPS proton measurements.** Here bin sizes are 3 hr for time, 1 hr for MLT and 0.2 for L-shell. **A1)** Binned fluxes of 16 MeV protons are presented as a function of L-shell and time for the 15-day period. **B1)** Distributions of 100 MeV protons presented in the same format. **C1, C2, & C3)** Proton distributions at the time bin of 148.5 hr with low flux levels: **C1** for L-MLT distribution for 16 MeV protons, **C2** for 100 MeV protons, and **C3** for energy-L distributions in the Northern hemisphere. **D1-D3)** Proton distributions at the time of 307.5 hr with moderate flux levels. **E1-E3)** Proton distributions for the time of 166.5 hr with high flux levels.

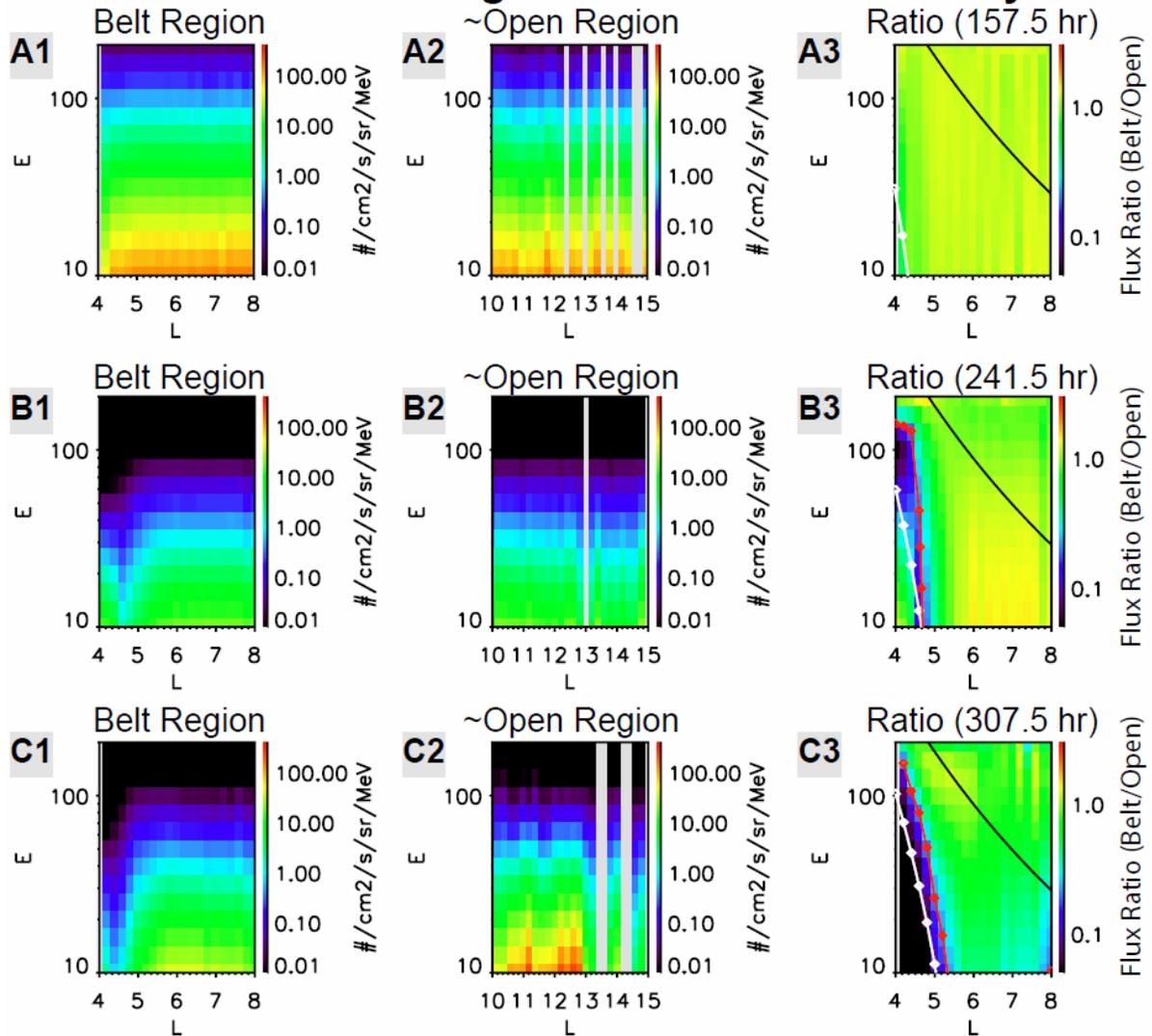

**Figure 9 Proton flux distributions and cutoff L-shells determined at three selected moments.** In the 1st row for the time bin of 157.5 hr in SPE1, **A1)** Flux distributions of protons inside the closed-field line region (L< =8, also called belt region), **A2)** flux distribution of protons in the open-field line region, and **A3)** ratios between fluxes inside the belt region and the open-field line region. The black line is cutoff L determined using a dipole magnetic field, and the white line is cutoff L determined from the empirical relation by Ogliore et al. (2001) and Leske et al. (2001). The red line is cutoff L determined by GPS observations. Panels in the 2nd row are for the time 241.5 hr (SPE1), and the 3rd row is for the time of 307.5 hr (SPE2).

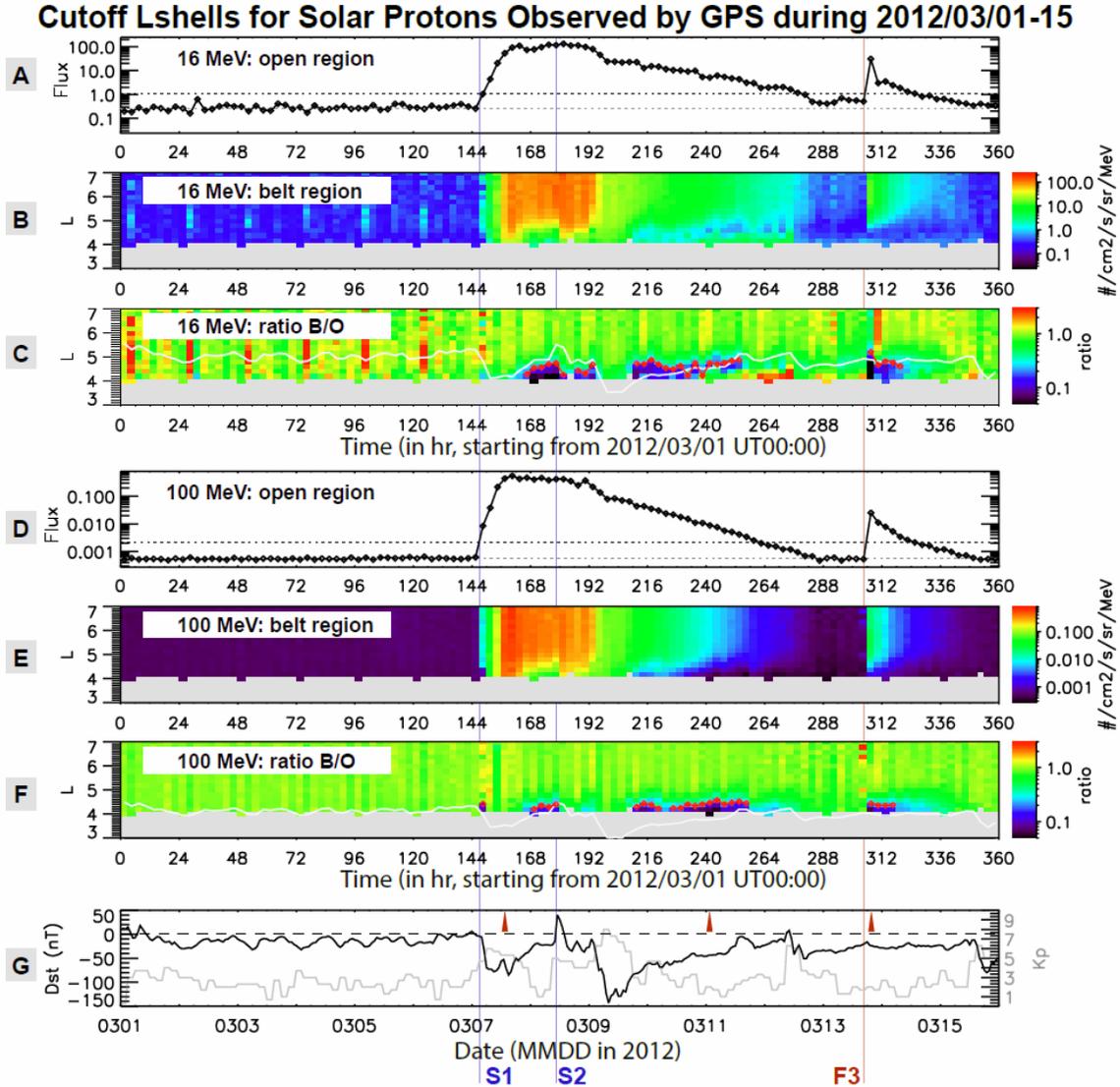

**Figure 10 Temporally evolving cutoff L-shells are identifiable from GPS data for the two SPEs.** The top three panels are for 16 MeV protons: **A)** Proton fluxes inside the open-field line region vary with time over the 15 day period. **B)** Proton distributions inside the belt region vary with time. **C)** Flux ratios between the belt region and open-field line region are used for identifying cutoff L-shells (red data points), compared to those determined from empirical relationship (in white). **D, E & F)** Flux distributions for 100 MeV protons in the same format as the top three panels. **G)** Dst (black) and Kp (gray) indices. Times for the shocks S1 and S2 as well as the solar flare F3 are marked out by vertical lines. The three red triangles indicate the three selected moments as in Figure 9.

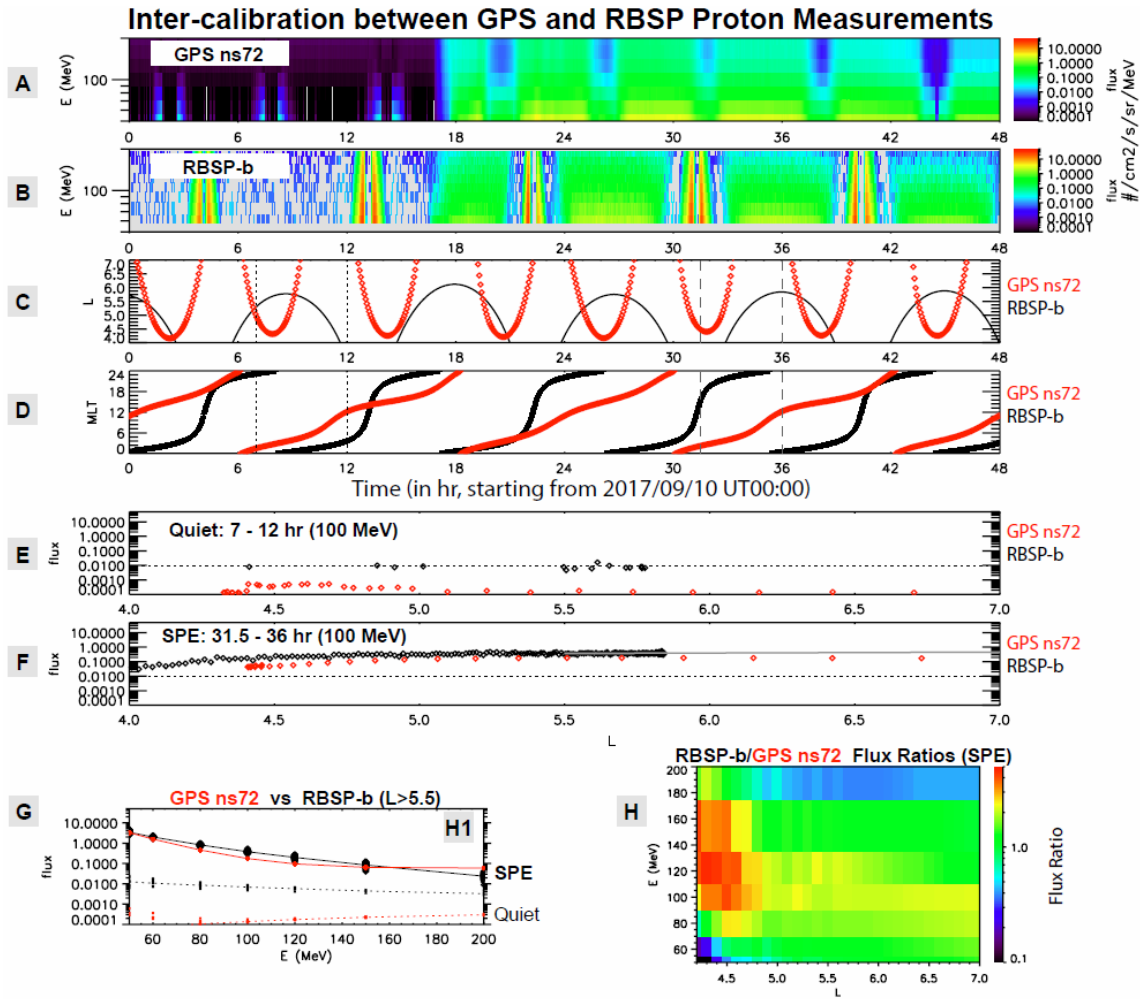

**Figure 11 Comparing proton measurements from GPS CXD and RBSP RPS instruments over two days in September 2017.** An SPE occurs starting from ~16.5 hr. **A)** Proton fluxes measured by GPS ns72 over two days 10-11 Sep. **B)** Proton fluxes measured by RBSP-b over the same period. **C)** L-shells of ns72 (in red) and RBSP-b (black). **D)** MLT of ns72 (red) and RBSP-b (black). **E)** 100 MeV proton fluxes from ns72 (red) compared to those from RBSP-b (black) during 5 hr quiet time as a function of L-shell. **F)** 100 MeV proton fluxes from ns72 (red) vs those of RBSP-b (black) during ~5 hr of SPE. The gray line extrapolates RBSP fluxes to higher L-shells. **G)** Proton fluxes at L > 5.5 as a function of energy during SPE (solid lines) and quiet time (dashed). Lines connect fluxes averaged over individual data points at each energy. **H)** Flux ratios (RBSP-b over ns72) averaged over the SPE hours as a function of L-shell and energy.

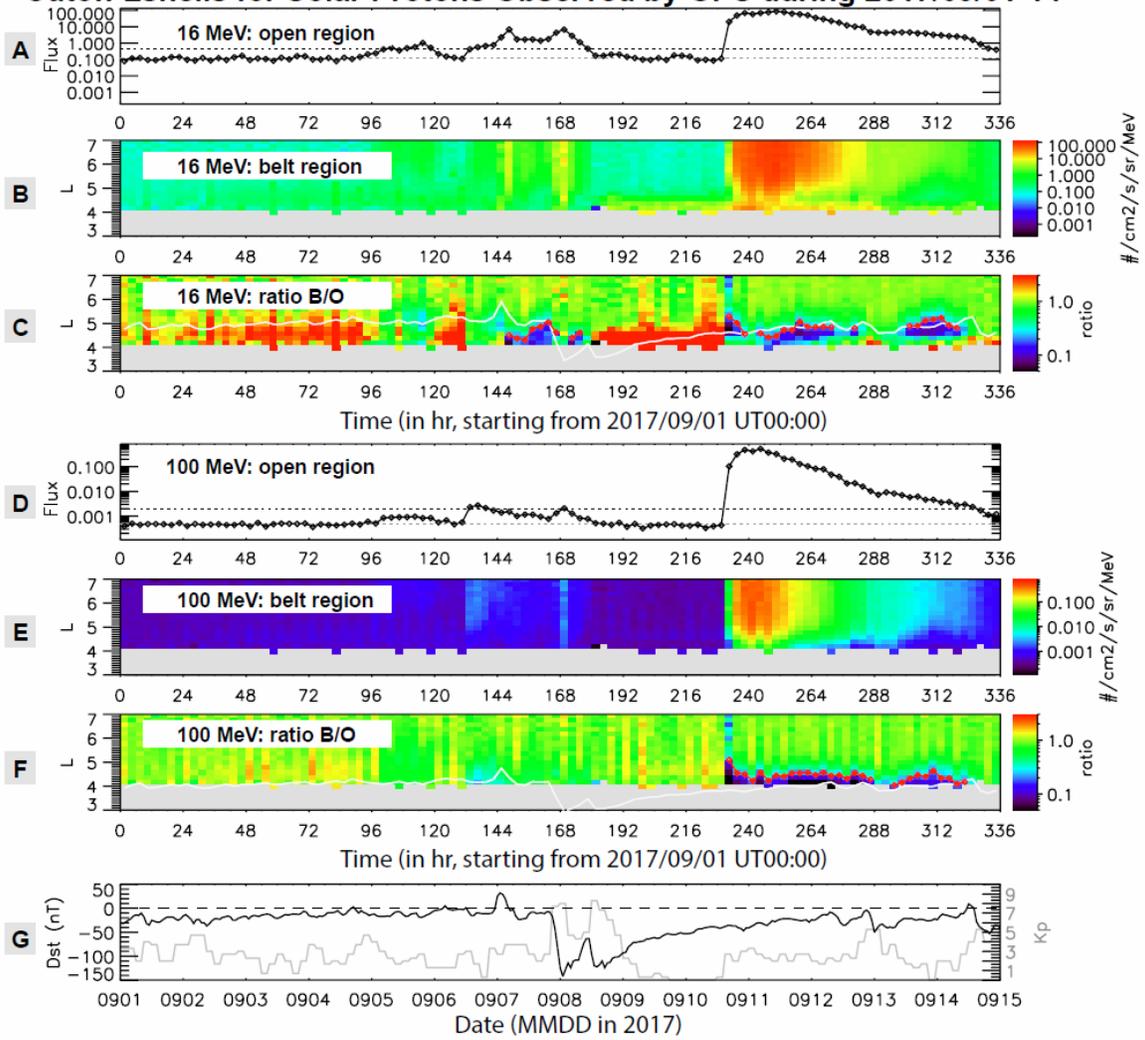

**Figure 12 Temporally evolving cutoff L-shells are identified for SPEs in September 2017.** Panels are in the same format as Figure 10.